\documentclass[twocolumn]{IEEEtran}

\usepackage{graphicx}

\usepackage{amsmath}

\usepackage{amsfonts}

\newtheorem{thm}{Theorem}
\newtheorem{lem}{Lemma}
\newtheorem{Def}{Definition}
\newtheorem{rem}{Remark}

\newtheorem{exam}{Example}

\begin{document}
%
\title{Dark Modes of Quantum Linear Systems\thanks{This work is supported by the Australian Research Council (DP130101658, FL110100020).}}
%
%
%

\author{Yu~Pan,
        Daoyi~Dong,~\IEEEmembership{Senior Member,~IEEE,}
        and~Ian~R.~Petersen,~\IEEEmembership{Fellow,~IEEE}
\thanks{Y. Pan is with the Institute of Cyber-Systems and Control, Zhejiang University, Hangzhou 310027, China. D. Dong and I. R. Petersen are with the School of Engineering and Information Technology, the University of New South Wales, Canberra 2600, ACT, Australia. e-mail: yu.pan.83.yp@gmail.com (Yu Pan), daoyidong@gmail.com (Daoyi Dong), i.r.petersen@gmail.com (Ian R. Petersen).}
}

\maketitle

\begin{abstract}
In this paper, we develop a direct method for the characterization of dark modes. The results can be used to construct a transformation that separates dark and bright modes, through the decomposition of system dynamics. We also study a synthesis problem by engineering the system-environment coupling and Hamiltonian engineering. We apply the theory to investigate an optomechanical dark mode.
\end{abstract}


%
\IEEEpeerreviewmaketitle

\section{introduction}\label{secintro}
%
%
%
%
\IEEEPARstart{A}{} major obstacle in quantum information processing is the coherent manipulation of fragile quantum information in the presence of environmental noise. The coherence of quantum systems will be lost if the systems are perturbed by environmental noise. This process of losing quantum coherence is commonly called decoherence. One way to counteract the decoherence effect is by engineering a decoherence-free (DF) subsystem \cite{Lidar03,Ganesan07,Ticozzi08,Ticozzi20092002,Ticozzi10,ZhangJ10,Naoki14,Naoki14X,Gough15,PanACC}. According to the DF linear quantum subsystem theory developed in \cite{Naoki14}, the DF modes are defined as the uncontrollable and unobservable modes of a quantum linear system. The linear DF modes can be obtained using the standard uncontrollable and unobservable decomposition of a linear system \cite{rugh93,Naoki14,Gough15}.

In this paper, the definition of dark modes is as follows:
\begin{Def}\label{def1}
For quantum linear systems, the dark modes are uncontrollable and unobservable modes that are defined on an arbitrarily given Hilbert space which is associated with a subsystem.
\end{Def}

According to the above definition, a dark mode $x_D$ can be characterized by the following dynamical equation
\begin{eqnarray}\label{gd}
dx(t)&=&d\left(
\begin{array}{c}
x_D(t)\\
\hline
x_B(t)\\
x_d(t)
\end{array}
\right)\nonumber\\
&=&\left(
\begin{array}{cc}
\hat{A}_{11}&0\\
\hline
0&\hat{A}_{22}\\
\end{array}
\right)x(t)dt+\left(
\begin{array}{c}
0\\
\hline
\hat{B}_2\\
\end{array}
\right)d\mathcal W(t),\nonumber\\
d\mathcal W_{out}(t)&=&(0\ \vline\ \hat{C}_2)x(t)dt+d\mathcal W(t),
\end{eqnarray}
where $(x_D,x_B)$ are defined on the arbitrarily given Hilbert subspace $\mathcal D$ which is associated with a subsystem. For simplicity, we use the same notation $\mathcal D$ when referring to this subsystem. $x_d$ is defined on a Hilbert subspace $\mathcal N$ (subsystem $\mathcal N$). As a result, the quantum linear system is defined on $\mathcal D\otimes\mathcal N$. $\mathcal W(t)$ is a noise process and $\mathcal W_{out}(t)$ is an output process of the system. Based on (\ref{gd}), we have
\begin{equation}\label{introd1}
dx_{D}=\hat{A}_{11}x_{D}dt,
\end{equation}
which implies that the dynamics of $x_D$ is decoupled from $x_B$, the environmental noise and the noisy subsystem $\mathcal N$. $x_B$ is called a bright mode if it interacts with the noisy subsystem $\mathcal N$ via $\hat{A}_{22}$. The above definition of dark mode is consistent with the literature. e.g., see \cite{DongC12,Wang12,Shkarin14}, in which the formation of a dark mode is used to achieve mediation between subsystems while being decoupled from a given noisy subsystem.

In this paper, we develop a direct method for the decomposition of the system dynamics as in (\ref{gd}), based on a suitable coordinate transformation. After preliminaries are presented in Section~\ref{secpreli}, a direct method to characterize dark modes is developed in Section~\ref{secdark}. Section~\ref{syn1} discusses the synthesis of dark modes. By engineering the system-environment coupling operator, we can remove the direct coupling of the dark modes to the noise and subsystem $\mathcal N$. Assisted by suitable Hamiltonian engineering, the indirect coupling can be eliminated as well and dark modes are generated. In order to illustrate the applications of the dark mode theory, in Section~\ref{examdark} we study an optomechanical system which relies on dark modes to function.

Notation: $A^T$ denotes the transpose of $A$. $A^\dagger$ is the Hermitian adjoint of $A$. $N(A)=\{v\ \vline\ Av=0\}$ is the kernel of $A$, and $R(A)=\{Av,\forall v\}$ is the range of $A$. $R^\perp(A)$ is the orthogonal complement to $R(A)$. $A^+$ denotes the  Moore-Penrose generalized inverse of $A$. $[X,Y]=XY-YX$. $0_n$ is an $n$-dimensional zero matrix, and $I_n$ is an $n$-dimensional identity matrix. $\emptyset$ is the empty set. $\Re(a),\Im(a)$ are the real and imaginary parts of a complex number $a$. $\mbox{i}$ is the imaginary unit. $\Sigma_n=\mbox{diag}\{\Sigma,\cdot\cdot\cdot,\Sigma\}$ is a block diagonal matrix containing $n$ two-dimensional matrices $\Sigma$ defined by $\Sigma=\left(\begin{array}{cc}0&1\\-1&0\end{array}
\right)$. $\delta_{ij}=0 $ if $i\neq j$, and $\delta_{ij}=1$ if $i=j$. Also, we use $\delta(\cdot)$ to denote the Dirac delta function. $\rho$ denotes a quantum state which is a Hermitian operator satisfying trace$(\rho)=1$ and $\rho\geq0$. We set $\hbar=1$.

\section{preliminaries}\label{secpreli}

\subsection{Heisenberg-picture Evolution and Quantum Stochastic Differential Equations}
The dynamics of a quantum system can be characterized by the evolution of a quantum state defined as $\rho(t)=U(t,t_0)\rho(0)U(t,t_0)$, where $t_0$ is the initial time and $U(t_0,t_0)=I$. $U(t,t_0)$ is the unitary operator which generates the quantum evolution \cite{hudson84,gardiner04}. The expectation of a system operator $X$ at the state $\rho$ is calculated by $\langle X\rangle_{\rho}=\mbox{trace}(X\rho)$. Accordingly, we can define the Heisenberg-picture evolution of the system operator $X(t)$ via the relation $\langle X(t)\rangle_{\rho(0)}=\langle X\rangle_{\rho(t)}$.

We consider an open quantum system coupled to the environment through $m$ inputs. The coupling operators associated with the $m$ inputs are given by $\{L_i,i=1,\cdot\cdot\cdot,m\}$. We use $H_0$ to denote the Hamiltonian of the open quantum system. $\{b_i(t),i=1,\cdot\cdot\cdot,m\}$ are the bosonic field annihilation operators defined on the $m$ input fields. The field operators satisfy the relation $[b_i(t),b_j^\dagger(s)]=\delta(t-s)$ for $i=j$ and $[b_i(t),b_j^\dagger(s)]=0$ for $i\neq j$. The quantum Wiener process is formally defined by $\tilde{B}_i(t)=\int_0^tb_i(s)ds$. As a consequence, $d\tilde{B}_i(t)=\tilde{B}_i(t+dt)-\tilde{B}_i(t)$ is the operator-valued quantum Ito increment. Introducing standard approximations which lead to the Markovian dynamics of the quantum system and using the quantum Ito calculus, we can obtain \cite{hudson84,gardiner04,wiseman09}
\begin{eqnarray}\label{lem1p}
dU(t,t_0)&=&\{\sum_{i=1}^m(b_i^\dag(t)L_i-L_i^\dag b_i(t))\nonumber\\
&-&(\frac{1}{2}\sum_{i=1}^{m}L_i^\dag L_i+\mbox{i}H_0)\}U(t,t_0)dt,\ t\geq t_0.
\end{eqnarray}
Based on (\ref{lem1p}), the Heisenberg-picture evolution of $X(t)$ is derived as the quantum stochastic differential equation (QSDE)
\begin{eqnarray}
&&dX(t)=-\mbox{i}[X(t),H_0(t)]+\sum_{i=1}^m\{L_i^\dagger(t) X(t)L_i(t)\nonumber\\
&-&\frac{1}{2}L_i^\dagger(t) L_i(t)X(t)-\frac{1}{2}X(t)L_i^\dagger(t) L_i(t)\nonumber\\
&+&d\tilde{B}^\dagger_i(t)[X(t),L_i(t)]+[L_i^\dagger(t),X(t)]d\tilde{B}_i(t)\},\label{nopr1}
\end{eqnarray}
with the input-output relation given by
\begin{eqnarray}\label{inout}
d\tilde{B}_{i,out}(t)&=&U^\dagger(t,t_0)d\tilde{B}_i(t)U(t,t_0)\nonumber\\
&=&L_i(t)+d\tilde{B}_i(t).
\end{eqnarray}
Here $\tilde{B}_{i,out}(t)=\int_0^tb_{i,out}(s)ds$ and $b_{i,out}(t)$ is the bosonic field annihilation operator defined on the $i$-th output.

\subsection{Quantum Linear Systems}\label{secIIB}
Quantum linear systems can be conveniently modelled using the QSDEs of system operators. In this paper, we consider a quantum linear system composed of $n$ harmonic oscillators, with $n_1$ oscillators defined on $\mathcal D$ and $n-n_1$ oscillators defined on $\mathcal N$. Each harmonic oscillator is identified by its position operator $x_i$ and momentum operator $p_i$. A linear combination of the operators $\{x_i,p_i\},i=1,\cdot\cdot\cdot,n$ is called a mode of the quantum system. The position and momentum operators satisfy the canonical commutation relation $[x_i,p_j]=\mbox{i}\hbar\delta_{ij}=\mbox{i}\delta_{ij}$. For convenience, the position and momentum operators are collected into a single vector as $x=(x_1,p_1,\cdot\cdot\cdot,x_n,p_n)^T$, which satisfies $xx^T-(xx^T)^T=\mbox{i}\Sigma_n$. Suppose that the system Hamiltonian $H_0$ and the system-environment coupling operators are given by
\begin{equation}
H_0=x^TGx,\quad L_i=c_i^Tx,\ i=1,2,\cdot\cdot\cdot,m,
\end{equation}
with $G$ being a $2n\times 2n$ real symmetric matrix and $c_i$ being a column vector of $2n$ scalars. According to (\ref{nopr1}), the dynamics of $x(t)$ are described by the following linear system equation
\begin{equation}\label{dflin}
dx(t)=Ax(t)dt+Bd\mathcal{W}(t).
\end{equation}
Here $\mathcal{W}=(X_1,P_1,\cdot\cdot\cdot,X_m,P_m)^T$ and $\{(X_i,P_i),i=1,\cdot\cdot\cdot,m\}$ are defined by $X_i=(\tilde{B}_i+\tilde{B}_i^\dag)/\sqrt{2},P_i=(\tilde{B}_i-\tilde{B}_i^\dag)/\sqrt{2}\mbox{i}$. $\mathcal{W}(t)$ is the noise process due to the environmental couplings. The coefficient matrices of (\ref{dflin}) are given by
\begin{eqnarray}
A&=&\Sigma_n(G+C^T\Sigma_mC/2)\in\mathbb{R}^{2n\times2n},\nonumber\\
B&=&\Sigma_nC^T\Sigma_m\in\mathbb{R}^{2n\times2m},
\end{eqnarray}
where the following definition is used \cite{Naoki14}
\begin{equation}\label{cmatrix}
C=\sqrt{2}(\Re(c_1),\Im(c_1),\cdot\cdot\cdot,\Re(c_m),\Im(c_m))^T\in\mathbb{R}^{2m\times2n}.
\end{equation}
Using (\ref{inout}), the input-output relation of the system can be written as
\begin{equation}\label{eqoutput}
d\mathcal W_{out}(t)=Cx(t)dt+d\mathcal W(t).
\end{equation}
A quantum linear system as expressed by (\ref{dflin}) and (\ref{eqoutput}) can be described by a triplet $(A,B,C)$.

A coordinate transformation $x=\mathcal Tx,\mathcal T\in\mathbb{R}^{2n\times2n}$ yields the following transformed system
\begin{eqnarray}\label{ct}
dx(t)&=&\mathcal T^{-1}A\mathcal Tx(t)dt+\mathcal T^{-1}Bd\mathcal W(t),\nonumber\\
d\mathcal W_{out}(t)&=&C\mathcal Tx(t)dt+d\mathcal W(t).
\end{eqnarray}

\subsection{Symplectic Matrices}
$\Sigma$ is a symplectic since $\Sigma^T\Sigma\Sigma=-\Sigma^T=\Sigma$. Consider the commutator $[x_i,p_i]$ as a bilinear form which can be expressed as $(x_i\ p_i)\Sigma(x_i\ p_i)^T$. Since $\Sigma^T\Sigma\Sigma=\Sigma$, $\Sigma(x_i\ p_i)^T$ is the symplectic transformation that preserves the commutation relation between the canonically conjugate operators of the harmonic oscillator. The same argument also applies to $\Sigma_n$. As a result, the symplectic matrix $\Sigma_n$ plays a fundamental role in the transformations of the physically realizable quantum systems, e.g. see \cite{souriau2012structure,HNurdin14,Tech15}. The transfer function of the passive symplectic system (\ref{ct}) obeys the relation $G(\mbox{i}\omega)^\dag G(\mbox{i}\omega)=G(\mbox{i}\omega)G(\mbox{i}\omega)^\dag=I_m$ for all $\omega\in \mathbb{R}$ \cite{Gough15}.

\section{characterization of dark modes}\label{secdark}
We will refer to the following result from \cite{Naoki14}.
\begin{lem}\label{lemma1}
$\Sigma_nv$ and $v$ are orthogonal column vectors, i.e. $(\Sigma_nv)^Tv=-v^T\Sigma_nv=0$. If $v$ is a normalized vector, i.e. $v^Tv=1$, then $\Sigma_nv$ is a normalized vector as well.
\end{lem}

Define a $\mathcal P$-matrix as
\begin{equation}\label{bcon}
\mathcal P_{m,n}(X)=\left(
\begin{array}{c}
\Sigma_{m}X\Sigma_{n}\\
X
\end{array}
\right),
\end{equation}
where $X$ is an arbitrary $2m\times2n$ matrix. $\mathcal P_{m,n}(X)$ is a $4m\times2n$ matrix. We can prove the following lemma.
\begin{lem}\label{lemma2}
Suppose a normalized vector $v$ is in $N(\mathcal P_{m,n}(X))$, i.e.
\begin{equation}\label{dcon}
\mathcal P_{m,n}(X)v=\left(
\begin{array}{c}
\Sigma_{m}X\Sigma_{n}\\
X
\end{array}
\right)v=0.
\end{equation}
Then $\Sigma_{n}v$ is a normalized vector in $N(\mathcal P_{m,n}(X))$ as well.
\end{lem}
\begin{IEEEproof}
$\Sigma_{n}v$ is normalized from Lemma~\ref{lemma1}. Using (\ref{dcon}) we have $\Sigma_{m}\Sigma_{m}X\Sigma_{n}v=0$. Hence, $-I_{m}X\Sigma_{n}v=0$. Therefore, $\Sigma_{m}X\Sigma_{n}(\Sigma_{n}v)=-\Sigma_{m}Xv=0$ and $X(\Sigma_{n}v)=0$
hold, which proves that $\Sigma_{n}v$ is also in the kernel of $\mathcal P_{m,n}(X)$.
\end{IEEEproof}

Lemma~\ref{lemma2} implies that if a vector $v$ is in the intersection of $N(B^T)$ and $N(C)$ ($v$ is decoupled from the direct interaction with the input and output), then its symplectic transformation $\Sigma_nv$ is also in $N(B^T)\cap N(C)$. This will result in a symplectic coordinate transformation matrix $\mathcal T$ in generating the dark modes.

As explained in Sec.~\ref{secIIB}, the vector $x(t)$ can be decomposed as $x(t)=(x_{\mathcal D}^T\quad x_{\mathcal N}^T)^T$, where $x_{\mathcal D}=(x_1\ p_1\cdot\cdot\cdot x_{n_1}\ p_{n_1})^T$ is the collection of the system operators for the $n_1$ harmonic oscillators that constitute the subsystem $\mathcal D$, and $x_{\mathcal N}=(x_{n_1+1}\ p_{n_1+1}\cdot\cdot\cdot x_{n}\ p_{n})^T$ is the collection of the system operators for the $n-n_1$ harmonic oscillators that constitute the subsystem $\mathcal N$. Accordingly, $C$ is decomposed as $C=(C_1\quad C_2),C_1\in\mathbb{R}^{2m\times 2n_1},C_2\in\mathbb{R}^{2m\times 2(n-n_1)}$. Note that $x_{\mathcal D}$ is directly coupled to the input $\mathcal W(t)$ via $B_1=\Sigma_{n_1}C_1^T\Sigma_m\in\mathbb{R}^{2n_1\times 2m}$, and to the output via $C_1$. The system Hamiltonian can be written as $H_0=H_{int}+H_{\mathcal D}+H_{\mathcal N}$ with $H_{int}=x^TG_{int}x$ being the interaction Hamiltonian between the two subsystems, and the internal Hamiltonians of the two subsystems are written as $H_{\mathcal D}=x^TG_{\mathcal D}x$ and $H_{\mathcal N}=x^TG_{\mathcal N}x$, respectively. We decompose $G_{int}\in\mathbb{R}^{2n\times 2n}$ as $G_{int}=\left(
\begin{array}{c}
G_{1,int}\\
G_{2,int}
\end{array}
\right)$ with $\ G_{1,int}\in\mathbb{R}^{2n_1\times 2n}$. Furthermore, for simplicity we denote $\mathcal P_{m+n,n_1}(\left(
\begin{array}{c}
C_1\\
G_{1,int}^T
\end{array}
\right))\in\mathbb{R}^{2(m+n)\times 2n_1}$ as $\mathcal P$. The rank defect of this $\mathcal P$ may indicate the existence of dark modes.

\begin{lem}\label{lemman3}
If rank$(\mathcal P)=q<2n_1$, then the system equations (\ref{dflin}) and (\ref{eqoutput}) can be transformed to
\begin{eqnarray}\label{darkeq}
&&d\left(
\begin{array}{c}
x_D(t)\\
\hline
x_B(t)\\
x_d(t)
\end{array}
\right)\nonumber\\
&=&\left(
\begin{array}{ccc}
P_1^T\Sigma_nGP_1&\vline& P_1^T\Sigma_nG_{\mathcal D}P_2\\
\hline
P_2^T\Sigma_nG_{\mathcal D}P_1&\vline& P_2^TAP_2
\end{array}
\right)x(t)dt\nonumber\\
&+&\left(
\begin{array}{c}
0\\
\hline
P_2^T \Sigma_nC^T\Sigma_m
\end{array}
\right)d\mathcal W(t),\nonumber\\
&&d\mathcal W_{out}(t)=(0\quad CP_2)x(t)dt+d\mathcal W(t),
\end{eqnarray}
under a proper coordinate transformation $x=\mathcal Tx=(P_1\quad P_2)x$, with $x_D$ containing at least two modes.
\end{lem}

\begin{IEEEproof}
We have rank$(N(\mathcal P))+$rank$(\mathcal P)=2n_1$ due to the rank-nullity theorem. If rank$(\mathcal P)=q<2n_1$, then rank$(N(\mathcal P))=2n_1-q>0$. Then we can construct a transformation matrix $\mathcal T=(P_1\quad P_2)$ by letting
\begin{equation}
P_1=\left(
\begin{array}{ccccc}
\cdot\cdot\cdot&v_i& \Sigma_{n_1}v_i&\cdot\cdot\cdot\\
0&0&0&0
\end{array}
\right)\in\mathbb{R}^{2n\times (2n_1-q)},
\end{equation}
with $\{v_i,\Sigma_{n_1}v_i,i=1,2,\cdot\cdot\cdot,(2n_1-q)/2,v_i\in\mathbb{R}^{2n_1\times 1}\}$ being the basis vectors of $N(\mathcal P)$. Here we have made use of Lemma~\ref{lemma2}. Note that the last $2(n-n_1)$ rows of $P_1$ are set as $0$. $v_i$ is chosen to be orthogonal to $v_j,\Sigma_{n_1}v_j$ for all $j<i$.  By this construction, the column vectors of $P_1$ are mutually orthogonal. Next, we construct $P_2\in\mathbb{R}^{2n\times(2n-2n_1+q)}$ as $P_2=(v_{(2n_1-q)/2+1}\ \Sigma_nv_{(2n_1-q)/2+1}\cdot\cdot\cdot v_n\ \Sigma_nv_n)$ which is composed of mutually-orthogonal normalized column vectors which are orthogonal to all the column vectors of $P_1$. Here we can choose $P_2$ to have this form based on Lemma~\ref{lemma1}. Then we can verify that $\mathcal T\in \mbox{Sp}(2n)\cap \mbox{O}(2n)$, where $\mbox{Sp}(2n)$ is the symplectic group of $2n\times2n$ matrices and $\mbox{O}(2n)$ is the orthogonal group of $2n\times2n$ matrices. The transformed coefficient matrices in (\ref{ct}) can be calculated using the relations $CP_1=0,B^TP_1=0,P_1^TG_{int}=0$ as well as $P_1^T\Sigma_nG_{int}=0$. Also note that $P_1^TG_{\mathcal N}=0$ and $P_1^T\Sigma_nG_{\mathcal N}=0$ are automatically satisfied because the last $2(n-n_1)$ rows of $P_1$ are zero. Therefore, we have $P_1^TAP_2=P_1^T\Sigma_nGP_2=P_1^T\Sigma_nG_{\mathcal D}P_2$ and $P_2^TAP_1=P_2^T\Sigma_nG_{\mathcal D}P_1$. The resulting system equations are thus given by (\ref{darkeq}).
\end{IEEEproof}

The modes $x_D$ in (\ref{darkeq}) are not directly coupled to the noise. However, $x_D$ may be indirectly coupled to the noise via $x_B$ and $x_d$. Therefore, dark modes can only be generated after we remove the coupling between $x_D$ and $x_B,x_d$. This can be done by engineering the Hamiltonian $H_{\mathcal D}$ of the subsystem $\mathcal D$. The following theorem provides a sufficient condition for the existence of dark modes.
\begin{thm}\label{theoremd1}
Suppose rank$(\mathcal P)=q<2n_1$ and the transformed system is given by (\ref{darkeq}). If the condition
\begin{equation}\label{condark1}
P_1^TG_{\mathcal D}P_2=0
\end{equation}
is satisfied, then $x_{D}$ are dark modes and the system equations become
\begin{eqnarray}\label{hc1}
d\left(
\begin{array}{c}
x_D(t)\\
\hline
x_B(t)\\
x_d(t)
\end{array}
\right)&=&\left(
\begin{array}{cc}
P_1^T\Sigma_nGP_1&0\\
\hline
0&P_2^TAP_2
\end{array}
\right)x(t)dt\nonumber\\
&+&\left(
\begin{array}{c}
0\\
\hline
P_2^T \Sigma_nC^T\Sigma_m
\end{array}
\right)d\mathcal W(t),\nonumber\\
d\mathcal W_{out}(t)&=&(0\quad CP_2)x(t)dt+d\mathcal W(t).
\end{eqnarray}
\end{thm}

\begin{IEEEproof}
According to Lemma~\ref{lemman3}, the condition (\ref{condark1}) can be explicitly written as
\begin{eqnarray}\label{thm1p1}
(v_i^T\ 0)G_{\mathcal D}v_{(2n_1-q)/2+j}&=&0,\nonumber\\
-(v_i^T\ 0)\Sigma_nG_{\mathcal D}v_{(2n_1-q)/2+j}&=&0,\nonumber\\
(v_i^T\ 0)G_{\mathcal D}\Sigma_nv_{(2n_1-q)/2+j}&=&0,\nonumber\\
-(v_i^T\ 0)\Sigma_nG_{\mathcal D}\Sigma_nv_{(2n_1-q)/2+j}&=&0,
\end{eqnarray}
for $i=1,\cdot\cdot\cdot,(2n_1-q)/2,\ j=1,\cdot\cdot\cdot,(2n-2n_1+q)/2$. The elements of the matrix $P_1^T\Sigma_nG_{\mathcal D}P_2$ are expressed as
\begin{eqnarray}
&&(v_i^T\ 0)\Sigma_nG_{\mathcal D}v_{(2n_1-q)/2+j},\nonumber\\
&&(v_i^T\ 0)G_{\mathcal D}v_{(2n_1-q)/2+j},\nonumber\\
&&(v_i^T\ 0)\Sigma_nG_{\mathcal D}\Sigma_nv_{(2n_1-q)/2+j},\nonumber\\
&&(v_i^T\ 0)G_{\mathcal D}\Sigma_nv_{(2n_1-q)/2+j},
\end{eqnarray}
for $i=1,\cdot\cdot\cdot,(2n_1-q)/2,\ j=1,\cdot\cdot\cdot,(2n-2n_1+q)/2$. Hence, we can conclude that $P_1^T\Sigma_nG_{\mathcal D}P_2=0$ by (\ref{thm1p1}). Similarly, we can prove $P_2^T\Sigma_nG_{\mathcal D}P_1=0$.
\end{IEEEproof}
Condition (\ref{condark1}) proposes a Hamiltonian engineering problem. Additionally, it is straightforward to identify the bright modes using (\ref{hc1}). $x_B$ contains $q$ modes which are linear combinations of the operators in $x_{\mathcal D}$. If these modes are coupled to the subsystem $\mathcal N$ via the interaction terms in $P_2^TAP_2$, then they are bright modes.

We can also consider the special case $C_1=0$. In this case, the sufficient conditions for the existence of dark modes are simplified as rank$(\mathcal P_{n,n_1}(G_{1,int}^T))<2n_1$ and (\ref{condark1}).

The dark modes are governed by the dynamical equation $\dot{x}_{D}=P_1^T\Sigma_nGP_1x_{D}$. If $P_1^T\Sigma_nGP_1=0$, then $\dot{x}_{D}=0$ and the dark modes are invariant. This fact can be summarized as the following theorem.
\begin{thm}\label{theorem2}
Suppose rank$(\mathcal P)=q<2n_1$ and the transformed system is given by (\ref{darkeq}). If the condition
\begin{equation}\label{sc}
P_1^TG_{\mathcal D}=0
\end{equation}
is satisfied, then the dark modes $x_{D}$ are invariant.
\end{thm}

\begin{rem}
The results of this section are closely related to the Popov-Belevitch-Hautus (PBH) controllability/observability criterion. Consider an equivalent statement of the PBH observability criterion: $(C,A)$ is unobservable if and only if there is a $v\neq0$ with $Av=\lambda v$ and $Cv=0$. Using $v\in N(C)$ we have $\Sigma_nGv=\lambda v$, which leads to $\lambda v^T=-v^TG\Sigma_n$. Similarly, if $v^TA=\mu v^T,v^TB=0$ for the same $v\neq0$, then the unobservable mode is also uncontrollable and we have $-v^T\Sigma_nG=\mu v^T$ using $v^TB=0$. Suppose $P_1,P_2$ are constructed using the same procedure as Lemma~\ref{lemman3}. Then we have $v^T\Sigma_nGP_2=-v^T\mu P_2=0$ and $P_2\Sigma_nGv=\lambda P_2v=0$ since $v$ is orthogonal to the column vectors of $P_2$. Using the coordinate transformation we can prove that the eigenvector $v$ corresponds to an uncontrollable and unobservable mode of the system. Here, $v\in N(B^T)\cap N(C)$ is equivalent to a rank-defect condition, and $\Sigma_nGv=\lambda v,-v^T\Sigma_nG=\mu v^T$ are conditions on the system Hamiltonian. So the PBH conditions combined with the direct method of this paper can be used to characterize linear DF modes. Furthermore, imposing the additional requirement that the dark mode is in the subsystem $\mathcal D$, then $C$ should be replaced with $C_1$ and the interaction between the dark mode and the subsystem $\mathcal N$ should be eliminated. Using this approach we will arrive at the sufficient conditions that are similar to the ones of Theorem~\ref{theoremd1}.

This connection to PBH criterion also suggests that the direct method of this paper can be used to characterize and engineer a mode that is only uncontrollable and unobservable from some specific inputs and outputs. The details of this application is presented in the Appendix.
\end{rem}

So far we have obtained a theory to characterize general dark modes. As shown in Theorem~\ref{theoremd1}, the existence of dark modes is conditioned in terms of the environmental couplings and the system Hamiltonian. In the next section, we consider the synthesis of dark modes through engineering the system-environment couplings followed by engineering the Hamiltonian.

\section{Engineering the system-environment couplings and Hamiltonian}\label{syn1}
Consider
\begin{equation}\label{darkgen}
\mathcal P_{m,n}(C)=\left(
\begin{array}{c}
\Sigma_mC\Sigma_n\\
C
\end{array}
\right),
\end{equation}
where $C\in\mathbb{R}^{2m\times2n}$ is the coefficient matrix for the environmental couplings associated with the subsystem $\mathcal D$. We assume that (\ref{darkgen}) is full column rank. As we have proven, no dark modes exist in this case.

Firstly, we demonstrate that adding couplings alone cannot reduce the column rank of the matrix. With the additional couplings, the coefficient matrix becomes
\begin{equation}
C^{'}=\left(
\begin{array}{c}
C\\
C_e
\end{array}
\right),
\end{equation}
where $C_e\in\mathbb{R}^{2m^{'}\times2n}$ is associated with the $m^{'}$ additional inputs. The updated $\mathcal P$-matrix for this system is thus $\mathcal P_{m+m^{'},n}(C^{'})$, the column rank of which is still $2n$ given that $\mathcal P_{m,n}(C)$ is full column rank.

For this reason, it is necessary to increase the dimension of the system. We consider three basic types of interconnections for increasing the dimension of the system, namely, cascade, direct coupling and coherent feedback \cite{PhysRevLett.70.2273,PhysRevLett.70.2269,Gough09,Tezak5270}.

\subsection{Cascade}
Suppose the original system and the additional system are defined by the triplets $(A_1,B_1,C_1)$ and $(A_2,B_2,C_2)$, respectively. Moreover, we assume that $C_1,C_2\in\mathbb{R}^{2m\times 2n}$. To form the cascade, the output of the original system is taken as the input to the additional system, which can be modelled as
\begin{eqnarray}
dx_i(t)&=&A_ix_i(t)dt+B_id\mathcal W_i(t),\quad i=1,2,\nonumber\\
d\mathcal W_{i,out}&=&C_ix_i(t)dt+d\mathcal W_i(t),\nonumber\\
d\mathcal W_2(t)&=&d\mathcal W_{1,out}.
\end{eqnarray}
The above equations can be rewritten as
\begin{eqnarray}\label{casexam}
dx_e(t)&=&\left(
\begin{array}{cc}
A_1&0\\
B_2C_1&A_2\\
\end{array}
\right)x_e(t)dt+\left(
\begin{array}{c}
B_1\\
B_2\\
\end{array}
\right)d\mathcal W_1(t),\nonumber\\
d\mathcal W_{2,out}&=&(C_1\quad C_2)x_e(t)dt+d\mathcal W_1(t),
\end{eqnarray}
where we have defined $x_e=(x_1\quad x_2)^T$. So we have
\begin{equation}
\mathcal P_{m,2n}((C_1\quad C_2))=\left(
\begin{array}{cc}
\Sigma_mC_1\Sigma_n&\Sigma_mC_2\Sigma_n\\
C_1&C_2\\
\end{array}
\right),
\end{equation}
whose column rank is $2n$ if we let $C_1=C_2$. Therefore, it is possible to generate $4n-2n=2n$ dark modes which are not influenced by the couplings associated with $(C_1\quad C_2)$.

It is easy to see that a weaker sufficient condition for $\mathcal P_{m,2n}((C_1\quad C_2))$ to be not full column rank is that at least one column vector of $\left(
\begin{array}{c}
\Sigma_mC_2\Sigma_n\\
C_2\\
\end{array}
\right)$ lies in the column space of $\left(
\begin{array}{c}
\Sigma_mC_1\Sigma_n\\
C_1\\
\end{array}
\right)$.

\subsection{Direct coupling}
The original system and the additional system are defined by the triplets $(A_1,B_1,C_1)$ and $(A_2,B_2,C_2)$, with $C_1,C_2\in\mathbb{R}^{2m\times 2n}$. The direct coupling is implemented by adding an interaction Hamiltonian $H_{int}=x^TG_{int}x$ between the two systems. If both $\mathcal P_{m,n}(C_1)$ and $\mathcal P_{m,n}(C_2)$ are full column rank, then we have
\begin{equation}
\mathcal P_{2m,2n}(\left(
\begin{array}{cc}
C_1&0\\
0&C_2
\end{array}
\right))=\left(
\begin{array}{cc}
\mathcal D_{m,n}(C_1)&0\\
0&\mathcal D_{m,n}(C_2)
\end{array}
\right),
\end{equation}
which is still full column rank. No dark modes exist in the augmented system.

\subsection{Coherent feedback}\label{seccf}
We consider two types of coherent feedback. The first type is modelled as
\begin{eqnarray}
dx_1(t)&=&A_1x_1(t)dt+B_1d\mathcal W_1(t)+B_2d\mathcal W_2(t),\nonumber\\
dx_2(t)&=&A_2x_2(t)dt+B_3d\mathcal W_3(t),\nonumber\\
d\mathcal W_{i,out}&=&C_ix_1(t)dt+d\mathcal W_i(t),\quad i=1,2,\nonumber\\
d\mathcal W_{3,out}&=&C_3x_2(t)dt+d\mathcal W_3(t),
\end{eqnarray}
where the original system and the additional system are defined by the triplets $(A_1,(B_1,B_2),(C_1,C_2)),A_1\in\mathbb{R}^{2n\times 2n},B_1,B_2\in\mathbb{R}^{2n\times 2m},C_1,C_2\in\mathbb{R}^{2m\times 2n}$ and $(A_2,B_3,C_3),A_2\in\mathbb{R}^{2n\times 2n},B_3\in\mathbb{R}^{2n\times 2m},C_3\in\mathbb{R}^{2m\times 2n}$, respectively. The additional system serves as the coherent controller, which processes the output $d\mathcal W_{1,out}(t)$ of the original system and feeds its output back to the original system. To close the loop we let $d\mathcal W_3=d\mathcal W_{1,out}$ and $d\mathcal W_2=d\mathcal W_{3,out}$. The closed-loop system is expressed as
\begin{eqnarray}
dx_e(t)&=&\left(
\begin{array}{cc}
A_1+B_2C_1&B_2C_3\\
B_3C_1&A_2\\
\end{array}
\right)x_e(t)dt\nonumber\\
&+&\left(
\begin{array}{c}
B_1+B_2\\
B_3\\
\end{array}
\right)d\mathcal W_1(t),\nonumber\\
d\mathcal W_{2,out}&=&(C_1+C_2\quad C_3)x_e(t)dt+d\mathcal W_1(t).
\end{eqnarray}
Similar to the cascade case, if we let $C_3=C_1+C_2$, then the matrix $\mathcal P_{m,2n}((C_1+C_2\quad C_3))$ for the closed-loop system does not have full column rank.

The second type of closed-loop system is the cross feedback between two systems. In this case, the additional system is defined by $(A_2,(B_3,B_4),(C_3,C_4)),B_3,B_4\in\mathbb{R}^{2n\times 2m},C_3,C_4\in\mathbb{R}^{2m\times 2n}$ with two inputs and two outputs. The system equations are given by
\begin{eqnarray}\label{cfbefore}
dx_1(t)&=&A_1x_1(t)dt+B_1d\mathcal W_1(t)+B_2d\mathcal W_2(t),\nonumber\\
dx_2(t)&=&A_2x_2(t)dt+B_3d\mathcal W_3(t)+B_4d\mathcal W_4(t),\nonumber\\
d\mathcal W_{i,out}&=&C_ix_1(t)dt+d\mathcal W_i(t),\quad i=1,2,\nonumber\\
d\mathcal W_{j,out}&=&C_jx_2(t)dt+d\mathcal W_j(t),\quad j=3,4.
\end{eqnarray}
The cross feedback is realized by letting $d\mathcal W_3=d\mathcal W_{1,out}$ and $d\mathcal W_2=d\mathcal W_{4,out}$, which transform the system equations to
\begin{eqnarray}\label{cfeq1}
dx_e(t)&=&\left(
\begin{array}{cc}
A_1&B_2C_4\\
B_3C_1&A_2\\
\end{array}
\right)x_e(t)dt\nonumber\\
&+&\left(
\begin{array}{cc}
B_1&B_2\\
B_3&B_4\\
\end{array}
\right)\left(
\begin{array}{c}
d\mathcal W_1(t)\\
d\mathcal W_4(t)\\
\end{array}
\right),\nonumber\\
d\mathcal W_{2,out}(t)&=&(C_2\quad C_4)x_e(t)dt+d\mathcal W_4(t),\nonumber\\
d\mathcal W_{3,out}(t)&=&(C_1\quad C_3)x_e(t)dt+d\mathcal W_1(t).
\end{eqnarray}
By (\ref{cfeq1}), if we let $C_3=C_1$ and $C_4=C_2$, the rank-defect condition of Theorem~\ref{theoremd1} is satisfied and dark modes may exist.

\begin{exam}
In this example we consider the cross feedback design using two linear systems. The system operators are denoted as $(x_i,p_i),i=1,2$, where $x_i,p_i$ are position and momentum operators of the harmonic oscillators, respectively. Each system has two inputs and two outputs. The four coupling operators $L_i=\sqrt{\kappa}(x_1+\mbox{i}p_1)/\sqrt{2},i=1,2,L_j=\sqrt{\kappa}(x_2+\mbox{i}p_2)/\sqrt{2},j=3,4$ have equal coupling strength. We can obtain the following coefficient matrices for (\ref{cfbefore}):
\begin{eqnarray}
&&B_1=B_2=B_3=B_4=-\sqrt{\kappa}\left(
\begin{array}{cc}
1&0\\
0&1\\
\end{array}
\right),\nonumber\\
&&C_1=C_2=C_3=C_4=-B_1^T.
\end{eqnarray}
Using the closed-loop equation (\ref{cfeq1}), it is straightforward to verify that $x_1-x_2$ and $p_1-p_2$ are dark modes which are decoupled from $L_1$ and $L_4$ if
\begin{equation}
P_1^T\left(
\begin{array}{cc}
A_1&B_2C_4\\
B_3C_1&A_2
\end{array}
\right)
P_2=0
\end{equation}
holds for
\begin{equation}
P_1=\frac{1}{\sqrt{2}}\left(
\begin{array}{cc}
I_2\\
-I_2
\end{array}
\right),\quad P_2=\frac{1}{\sqrt{2}}\left(
\begin{array}{cc}
I_2\\
I_2
\end{array}
\right).
\end{equation}
This condition leads to
\begin{equation}\label{robusta}
A_1=A_2,
\end{equation}
or $G_1=G_2$. Therefore, a sufficient condition for the existence of dark modes is that the Hamiltonians of the two systems are the same. Eq.~(\ref{robusta}) shows that cross feedback provides a robust realization for dark modes. The structures and parameters of the two linear systems can be uncertain, as long as two identical systems can be fabricated for cross feedback. Also, it is worth mentioning that in this case we have generated the dark modes under the condition that each system is coupled to full-rank noises $\{B_i\}$.

\end{exam}

\subsection{Hamiltonian engineering}

We consider the solution $G$ to the Hamiltonian engineering problems $P_1^TGP_2=0$ and $P_1^TG=0$ for the generation of dark modes and invariant modes.

First, $G=I$ and $G=0$ are special solutions to $P_1^TGP_2=0$, and $G=0$ is a solution to $P_1^TG=0$. Also we have the following result.

\begin{thm}\label{theorem3}
If $R(P_2)$ is invariant under $G$, then $G$ is a solution to $P_1^TGP_2=0$. The general solution to $P_1^TG=0$ is given by
\begin{equation}
G=(I-P_1P_1^T)Z(I-P_1P_1^T),
\end{equation}
with $Z$ being an arbitrary matrix.
\end{thm}

\begin{IEEEproof}
The fact that $R(P_2)$ is invariant under $G$ implies $GP_2=P_2M$ for a matrix $M$. Then $P_1^TGP_2=0$ follows since $P_1^TP_2=0$.

The general solution to $P_1^TG=0$ is given by
\begin{equation}
G=(I-(P_1^T)^+P_1^T)Z(I-(P_1^T)^+P_1^T),
\end{equation}
with $Z$ being an arbitrary matrix. It is easy to verify that $(P_1^T)^+=P_1$.
\end{IEEEproof}

\section{application to the dark modes of a quantum optomechanical system}\label{examdark}
In this section, we apply our theoretical results to the analysis of optomechanical dark modes and bright modes. Optomechanical systems are conventionally used for the fundamental study of light-matter interaction. Recently, they have also found applications in quantum storage \cite{Zhang2015}. As mentioned in Section~\ref{secintro}, dark modes and bright modes play vital roles in these applications.

\begin{figure}
\includegraphics[scale=0.4]{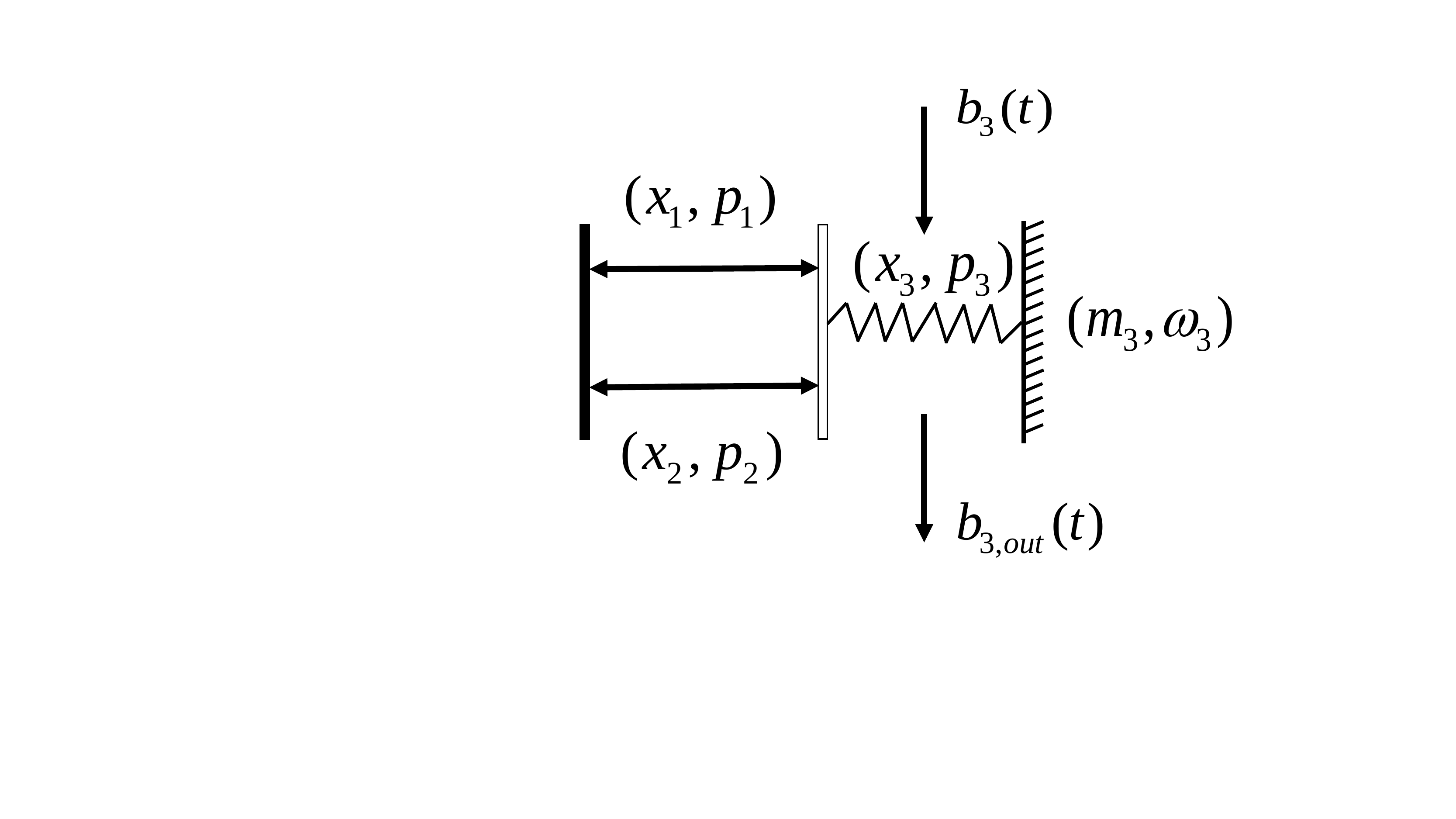}
\caption{An optomechanical dark mode. The position and momentum operators of the two optical modes are $(x_1,p_1)$ and $(x_2,p_2)$. The mechanical oscillator is characterized by $(x_3,p_3)$. The mechanical oscillator is coupled to the environment due to thermal dissipation.}
\label{fig1}
\end{figure}

We consider an optomechanical system which has been experimentally realized in \cite{DongC12}. The system is depicted in Fig.~\ref{fig1}. The two optical modes within optical cavities are coupled to a mechanical oscillator via the radiation pressure of the optical fields. The optical modes are modelled as harmonic oscillators with the system variables $(x_1,p_1)$ and $(x_2,p_2)$, respectively. $m_3$ and $\omega_3$ are the mass and frequency of the mechanical oscillator, and the mechanical mode is characterized by the variables $x_3$ and $p_3$. The mechanical oscillator is subjected to thermal noise, which can be modelled by a coupling operator $L=\sqrt{\kappa}b_3=\sqrt{\kappa}(x_3+\mbox{i}p_3)/\sqrt{2}$. $\kappa$ is the coupling strength. Since the direct coupling strength between the cavity and the environment is relatively small, we can ignore the corresponding optical losses.

This optomechanical setup is proposed to mediate the coupling between the two optical modes using the mechanical oscillator. However, the mechanical damping will undermine the mediated coupling and cause losses to the optical modes. One way to solve this problem is to exploit the optomechanical dark modes \cite{DongC12}. The dark modes are decoupled from the mechanical oscillator. The dark modes are the superposition of different optical modes. Therefore, the formation of stable dark modes is the result of the interaction and energy transfer between the optical fields. This process can be used to mediate an effective coupling between the optical modes and at the same time minimize the losses \cite{DongC12}. Another way to mediate coupling is to switch between dark and bright modes. Here, the bright mode is used to capture the photons, while the dark mode is used to store them.

For the proposed optomechanical system, we have
\begin{equation}
C=\sqrt{\kappa}(0_2\quad 0_2\quad I_2).
\end{equation}
Hence, we have $C_1=\sqrt{\kappa}(0_2\quad 0_2)$ for the engineering of dark modes that consist of optical modes only. The interaction Hamiltonian between the cavity and the mechanical oscillator is given by $H_{int}=\lambda_1a_1^\dag a_1x_3+\lambda_2a_2^\dag a_2x_3$, with $a_i=(x_i+\mbox{i}p_i)/\sqrt{2},i=1,2$. Applying a standard linearization procedure \cite{Aspel14} to $H_{int}$, we can obtain $G_{int}$ as
\begin{equation}
\begin{pmatrix}
  0_2 & 0_2& \begin{array}{cc}\gamma_1 & 0\\ 0 & 0\end{array} \\
  \hline
  0_2 & 0_2 & \begin{array}{cc} \gamma_2 & 0\\ 0 & 0\end{array} \\
  \hline
  \begin{array}{cc}\gamma_1 & 0\\ 0 & 0\end{array} &  \begin{array}{cc}\gamma_2 & 0\\ 0 & 0\end{array} & 0_2
 \end{pmatrix},
\end{equation}
where $\gamma_i,i=1,2$ is determined by $\kappa$ and $\lambda_i$. The $\mathcal P$-matrix as defined in Theorem~\ref{theoremd1} is given by
\begin{equation}
\begin{pmatrix}
0_4\\
\hline
0_4\\
\begin{array}{cccc}
\gamma_1&0&\gamma_2&0\\
0&0&0&0
\end{array}\\
\hline
0_4\\
\begin{array}{cccc}
0&0&0&0\\
0&-\gamma_1&0&-\gamma_2
\end{array}
\end{pmatrix},
\end{equation}
which is not full column rank. Therefore we can obtain
\begin{equation}
P_1=\left(
\begin{array}{cc}
a&0\\
0&a\\
b&0\\
0&b\\
0&0\\
0&0
\end{array}
\right),\quad P_2=\left(
\begin{array}{cc}
eI_2&0_2\\
fI_2&0_2\\
0_2&I_2
\end{array}
\right),
\end{equation}
with the condition $a\gamma_1+b\gamma_2=0$. $(e\quad f)^T$ and $(a\quad b)^T$ are required to be orthogonal. We then choose $G_{\mathcal D}$ such that $P_1^TG_{\mathcal D}P_2=0$. Finally, we can obtain the optomechanical dark modes as $ax_1+bx_2=\frac{\gamma_2}{\sqrt{\gamma_1^2+\gamma_2^2}}x_1-\frac{\gamma_1}{\sqrt{\gamma_1^2+\gamma_2^2}}x_2$ and its canonical conjugate mode $\frac{\gamma_2}{\sqrt{\gamma_1^2+\gamma_2^2}}p_1-\frac{\gamma_1}{\sqrt{\gamma_1^2+\gamma_2^2}}p_2$. It has been experimentally verified in \cite{DongC12} that these two modes are decoupled from the mechanical dissipation.

Note that any $G_{\mathcal D}$ satisfying $P_1^TG_{\mathcal D}P_2=0$ will generate the above dark modes. Here we assume a specific realization of the Hamiltonian $H_{\mathcal D}$ of the optical harmonic oscillators as $H_{\mathcal D}=\frac{p_1^2}{m_1}+m_1\omega_1^2x_1^2+\frac{p_2^2}{m_2}+m_2\omega_2^2x_2^2$. It is straightforward to verify that $P_1^TG_{\mathcal D}P_2=0$ if and only if $m_1=m_2$ and $\omega_1=\omega_2$. In other words, the dark modes exist if the two optical modes have the same energy. Experimentally, this can be realized by driving the optical modes with different frequencies $\omega_{l1}=\omega_{c1}-\omega_m$ and $\omega_{l2}=\omega_{c2}-\omega_m$, where $\{\omega_{ci},i=1,2\}$ are the cavity resonance frequencies.

Using (\ref{hc1}), we have $dx_D(t)=P_1^T\Sigma_nGP_1x_D(t)dt$ and
\begin{eqnarray}
d\left(
\begin{array}{c}
x_B(t)\\
x_d(t)
\end{array}
\right)&=&\left(
\begin{array}{cc}
0&P_2^TAP_2
\end{array}
\right)x(t)dt\nonumber\\
&+&P_2^T \Sigma_nC^T\Sigma_md\mathcal W(t),
\end{eqnarray}
with
\begin{eqnarray}\label{bmexam1}
&&P_2^TAP_2\nonumber\\
&&=\left(
\begin{array}{cccc}
0&\frac{1}{m_1}&0&0\\
-m_1\omega_1^2&0&-e\gamma_1-f\gamma_2&0\\
0&0&-\frac{\kappa}{2}&\frac{1}{m_3}\\
-e\gamma_1-f\gamma_2&0&-m_3\omega_3^2&-\frac{\kappa}{2}
\end{array}
\right),\nonumber\\
&&P_2^T \Sigma_nC^T\Sigma_m=-\sqrt{\kappa}\left(
\begin{array}{c}
0_2,\\
I_2
\end{array}
\right),
\end{eqnarray}
where we have $e=\frac{\gamma_1}{\sqrt{\gamma_1^2+\gamma_2^2}}$ and $f=\frac{\gamma_2}{\sqrt{\gamma_1^2+\gamma_2^2}}$. According to (\ref{bmexam1}), the bright mode $ep_1+fp_2$ is directly coupled to the mechanical mode, while $ex_1+fx_2$ is an indirect-coupled bright optical mode.

\section{Conclusion}
We have developed a direct method for the characterization and synthesis of dark modes. The key is to ensure that the interaction Hamiltonian between the subsystems does not affect the dark modes. Sufficient conditions are derived in terms of environmental couplings and the system Hamiltonian, which provides a straightforward and tractable way to engineer dark modes.

\appendix
\section*{Uncontrollable and unobservable modes from a specific input and output}\label{dfapp}
In order to decouple modes from a specific input and output, we may decompose $C$ as $C=(C_1^T\quad C_2^T)^T$. Here $C_1\in\mathbb{R}^{2n_1\times2n}$ is associated with $n_1$ coupling operators from $n_1$ inputs, and $C_2\in\mathbb{R}^{2(m-n_1)\times2n}$ is associated with the other $m-n_1$ inputs. The modes are required to decouple from the given $n_1$ inputs and outputs. If rank$(\left(
\begin{array}{c}
\Sigma_{n_1}C_1\Sigma_n\\
C_1
\end{array}
\right))=q<2n$, we can construct a transformation matrix $\mathcal T=(P_1\quad P_2)\in\mathbb{R}^{2n\times2n}$, where $P_1\in\mathbb{R}^{2n\times(2n-q)}$ is chosen as $P_1=(v_1\ \Sigma_nv_1\cdot\cdot\cdot v_{(2n-q)/2}\ \Sigma_nv_{(2n-q)/2})$ with $\{v_i,\Sigma_nv_i\}$ being mutually-orthogonal basis vectors of $N(\left(
\begin{array}{c}
\Sigma_{n_1}C_1\Sigma_n\\
C_1
\end{array}
\right))$. The $q$ column vectors of $P_2\in\mathbb{R}^{2n\times q}$ are chosen to be mutually-orthogonal normalized vectors which are orthogonal to the column vectors of $P_1$. The coordinate transformation yields
\begin{eqnarray}\label{darkcn1}
&&d\left(
\begin{array}{c}
x_{d1}\\
x_{d2}
\end{array}
\right)=\left(
\begin{array}{cc}
P_1^TAP_1&P_1^TAP_2\\
P_2^TAP_1&P_2^TAP_2
\end{array}
\right)xdt\nonumber\\
&&+\left(
\begin{array}{cc}
0&P_1^T\Sigma_nC_2^T\Sigma_{m-n_1}\\
P_2^T \Sigma_nC_1^T\Sigma_{n_1}&P_2^T\Sigma_nC_2^T\Sigma_{m-n_1}
\end{array}
\right)d\mathcal W,\nonumber\\
&&d\mathcal W_{out}=\left(
\begin{array}{cc}
0&C_1P_2\\
C_2P_1&C_2P_2
\end{array}
\right)xdt+d\mathcal W.
\end{eqnarray}

\begin{thm}
A sufficient condition for the modes $x_{d1}$ to be decoupled from the $n_1$ inputs is $P_1^TAP_2=0$.
\end{thm}

\begin{IEEEproof}
First, we decompose $d\mathcal W$ and $d\mathcal W_{out}$ as $(d\mathcal W_{d1}^T\quad d\mathcal W_{d2}^T)^T$ and $(d\mathcal W_{d1,out}^T\quad d\mathcal W_{d2,out}^T)^T$, where $\mathcal W_{d1}$ and $\mathcal W_{d1,out}$ are associated with $C_1$. Using this decomposition we have
\begin{eqnarray}\label{thm3eq1}
&&d{x_{d1}}=P_1^TAP_1x_{d1}dt+P_1^T\Sigma_nC_2^T\Sigma_{m-n_1}d\mathcal W_{d2},\nonumber\\
&&d\mathcal W_{d1,out}=C_1P_2x_ddt+d\mathcal W_{d1},
\end{eqnarray}
given that $P_1^TAP_2=0$ is satisfied. According to (\ref{thm3eq1}), $x_{d1}$ are decoupled from the input $\mathcal W_{d1}$, and the corresponding output $\mathcal W_{d1,out}$ is decoupled from $x_{d1}$ as well. Thus we have proven that $x_{d1}$ is neither controllable nor observable from the $n_1$ inputs.
\end{IEEEproof}


%





\ifCLASSOPTIONcaptionsoff
  \newpage
\fi











\end{document}